# Dopant-modulated lattice softening drives drastic thermal conductivity reduction in β-$FeSi_2$ thermoelectrics

Cuiping Zhang[1,2,3,4], Qingyong Ren[2,3,4*], Yangfan Cui[2,3,4], Chen Chen[5], Songbai Hu[5], Shengnan Dai[6], Chin-Wei Wang[7], Wanju Luo[2,3], Dexiang Gao[2,3], Bao Yuan[2,3], Junying Shen[2,3], Fan Chen[2,3], Wei Xu[1], Yuting Li[1], Mingfang Shu[1,8], Xiaoli Huang[9], Pengfei Qiu[10*], Jie Ma[1*]

[1]Key Laboratory of Artificial Structures and Quantum Control, School of Physics and Astronomy, Shanghai Jiao Tong University, Shanghai 200240, China

[2]Institute of High Energy Physics, Chinese Academy of Sciences, Beijing 100049, China

[3]Spallation Neutron Source Science Center, Dongguan 523803, China

[4]Guangdong Provincial Key Laboratory of Extreme Conditions, Dongguan 523803, China

[5]School of Physical Sciences, Great Bay University, Dongguan, Guangdong 523000, China

[6]Materials Genome Institute, State Key Laboratory of Advanced Refractories, Shanghai University, Shanghai, 200444, China.

[7]Neutron Group, National Synchrotron Radiation Research Center, Hsinchu 300092, Taiwan

[8]College of Sciences, China Jiliang University, Hangzhou 310018, China

[9]State Key Laboratory of High Pressure and Superhard Materials, College of Physics, Jilin University, Changchun 130012, China.

[10]State Key Laboratory of High Performance Ceramics, Shanghai Institute of Ceramics, Chinese Academy of Sciences, Shanghai 200050, China.

## Abstract:

Suppressing lattice thermal conductivity ($\kappa_{\text{lat}}$) is pivotal for thermoelectric efficiency. While traditional strategies rely heavily on phonon scattering from mass- and size-mismatches, we demonstrate a robust $\kappa_{\text{lat}}$ suppression mechanism driven by dopant-induced lattice stiffness modulation. Through a comparative analysis of p-type (Mn) and n-type (Co, Ir) doping in the β-$FeSi_2$ model system, we show that Co and Ir doping significantly reduce $\kappa_{\text{lat}}$. Notably, Co doping achieves a ~71% reduction at 300 K even without significant mass and size contrast. By correlating transport data with neutron powder diffraction, heat capacity, and Raman spectroscopy, we reveal anomalous lattice expansion, a substantial reduction in Debye temperature, and marked vibrational redshift and broadening. These systematic changes provide strong evidence for atomic-scale lattice softening and a fundamental weakening of interatomic force constants, which synergistically lower phonon group velocities and amplify anharmonic scattering. Our findings establish lattice stiffness manipulation as a powerful strategy for thermal management, offering a distinct design pathway beyond traditional mass- and strain-fluctuation models.

*Corresponding authors: Qingyong Ren, Pengfei Qiu, Jie Ma

Email: renqy@ihep.ac.cn; qiupf@mail.sic.ac.cn; jma3@sjtu.edu.cn

## 1. Introduction

Thermoelectric technology enables direct conversion between thermal and electrical energy, offering a key pathway for sustainable power generation and waste heat recovery[1]. The performance of thermoelectric materials is gauged by the dimensionless figure of merit, $ZT = S^2\sigma T/(\kappa_{\mathrm{ele}} + \kappa_{\mathrm{lat}})$, where $S$, $\sigma$, $T$, $\kappa_{\mathrm{ele}}$ and $\kappa_{\mathrm{lat}}$ denote the Seebeck coefficient, electrical conductivity, absolute temperature, electronic and lattice thermal conductivity, respectively[2-4]. Optimizing $ZT$ is challenging due to the strong interdependence of $\sigma$, $\kappa_{\mathrm{ele}}$, and $S$ [4,5]. Consequently, minimizing $\kappa_{\mathrm{lat}}$, which is governed by phonon transport and relatively independent of electronic properties, has become a primary strategy for performance enhancement[6-12]. Common approaches involve introducing lattice imperfections, such as point defects and grain boundaries, to enhance phonon scattering[13-18].

Traditional point defect engineering typically exploits mass and strain fluctuations to impede phonon transport[8,19,20]. However, recent studies suggest that specific dopants can influence lattice dynamics beyond simple scattering mechanisms[21-24]. Theoretical studies predict that doping can alter chemical bond strengths, inducing lattice softening and phonon spectrum reorganization, a mechanism distinct from the traditional mass/strain model[25]. Furthermore, experimental studies on multi-element doped GeTe-based high-entropy compounds revealed that additional local chemical fluctuations are essential to explain anomalous $\kappa_{\mathrm{lat}}$ reduction behavior[26,27]. Similarly, aliovalent doping in NbFeSb-based half Heusler compounds has been shown to decelerate high-frequency longitudinal optical phonons through dielectric screening and induce an avoided crossing between acoustic and low-lying optical branches, drastically reducing $\kappa_{\mathrm{lat}}$ [28,29]. These findings highlight that tailoring $\kappa_{\mathrm{lat}}$ requires understanding how dopants manipulate intrinsic lattice dynamics. Yet, direct experimental correlation between specific dopants and lattice stiffness modulation remains limited, hindering rational materials design.

Among numerous thermoelectric materials, β-$FeSi_2$ is a promising candidate for mid-to-high temperature applications due to its environmental friendliness, elemental abundance, and excellent thermal stability[30-39]. Moreover, the low expense of reaction elements, Fe and Si, offers it with an exceptional opportunity for the industrial products. However, its high intrinsic lattice thermal conductivity limits its thermoelectric performance[35]. Substitutional doping with transition metals (e.g., Co[35,40-42], Ir[32], Ru[31], Mn[33,43-45], Zr[34]) has been widely employed to suppress $\kappa_{\mathrm{lat}}$. Despite these efforts, suppression efficiencies vary significantly among different dopants, and the microscopic mechanisms driving these disparities, especially for dopants with similar atomic mass, remain poorly understood.

In this work, through a comprehensive investigation into the effects of Ir, Co, and Mn doping on the crystal structure, lattice dynamics, and thermal transport of the β-$FeSi_2$ model system, we unravel a missing link between specific dopants and lattice stiffness modulation. Unlike previous studies focusing on mass/size contrast or dielectric screening, we highlight the exceptional $\kappa_{\mathrm{lat}}$ suppression in Co-doped samples, which occurs despite negligible mass and size contrast with Fe. We show that this reduction correlates with a redshift in Raman frequencies. By integrating transport properties with heat capacity, neutron powder diffraction (NPD), and Raman spectroscopy, we demonstrate that the $\kappa_{\mathrm{lat}}$ suppression arises primarily from dopant-induced lattice softening and weakened interatomic force constants. This provides strong experimental evidence for a direct link between atomic-scale doping, intrinsic lattice stiffness, and macroscopic thermal transport, offering guiding principles for selecting dopants to achieve superior thermoelectric performance by design.

## 2. Results and Discussion

### 2.1 Thermoelectric Transport Properties

The phase purity and crystallinity of the synthesized samples were first examined through X-ray diffraction (XRD) at room temperature. As shown in Supplementary Fig. 1, all samples, $FeSi_2$, $(Fe,Mn)Si_2$, $(Fe,Co)Si_2$, and $(Fe,Ir)Si_2$, exhibit sharp peaks indexed to the orthorhombic β-phase structure. No secondary phases were detected, confirming high sample quality. The consistent diffraction patterns across all samples suggest successful dopant incorporation into the lattice without disrupting the fundamental crystal structure.

Fig. 1a presents the temperature-dependent Seebeck coefficient ($S$). The pristine $FeSi_2$ shows typical intrinsic semiconductor behavior with a low absolute $S$ and a thermally driven p-type to n-type crossover. Doping significantly alters this transport behavior. Mn doping yields a large positive $S$ (up to ~455 $\mu$V $K^{-1}$), confirming p-type conduction. In contrast, both Co and Ir doping induce n-type transport with negative $S$ values of ~-161 $\mu$V $K^{-1}$ and ~-84 $\mu$V $K^{-1}$, respectively. Simultaneously, the electrical conductivity ($\sigma$) increases by orders of magnitude (Fig. 1b). It is evident that the n-type dopants (Co, Ir) are more effective than Mn in enhancing carrier concentration (inset of Fig. 1c), consistent with previous reports[32].

The impact of doping on total thermal conductivity ($\kappa_{\mathrm{tot}}$) is shown in Fig. 1c. While Mn doping yields only a marginal reduction, Co and Ir doping significantly suppress $\kappa_{\mathrm{tot}}$. We isolate the lattice contribution ($\kappa_{\mathrm{lat}}$) by subtracting the electronic component ($\kappa_{\mathrm{ele}}$) using the Wiedemann-Franz Law ($\kappa_{\mathrm{ele}} = L\sigma T$, see Supplementary Fig. 2a for Lorenz number details). Despite the enhanced conductivity in doped samples, $\kappa_{\mathrm{ele}}$ remains a minor fraction of the total heat transport (Supplementary Fig. 2b).

Fig. 1d illustrates the lattice thermal conductivity $\kappa_{\mathrm{lat}}$. The pristine $FeSi_2$ displays a characteristic crystalline peak at low temperature (~50 K), signifying the dominance of Umklapp scattering. This feature persists in $(Fe,Mn)Si_2$, suggesting that Mn doping preserves the fundamental phonon scattering mechanisms. In sharp contrast, Co and Ir doping flattens this peak, leading to a “glass-like” temperature dependence. This points to the emergence of additional potent mechanisms that impede phonon propagation, leading to a much more pronounced $\kappa_{\mathrm{lat}}$ reduction. For instance, at 300 K, $\kappa_{\mathrm{lat}}$ drops to ~4.7 W $m^{-1}$ $K^{-1}$ for $(Fe,Co)Si_2$ and ~1.7 W $m^{-1}$ $K^{-1}$ for $(Fe,Ir)Si_2$, a drastic reduction compared to the ~21.9 W $m^{-1}$ $K^{-1}$ of the Mn-doped sample and ~27.0 W $m^{-1}$ $K^{-1}$ of the pristine host, or representing a reduction of ~71%, ~88%, and 19% for $(Fe,Co)Si_2$, $(Fe,Ir)Si_2$, and $(Fe,Mn)Si_2$, respectively.

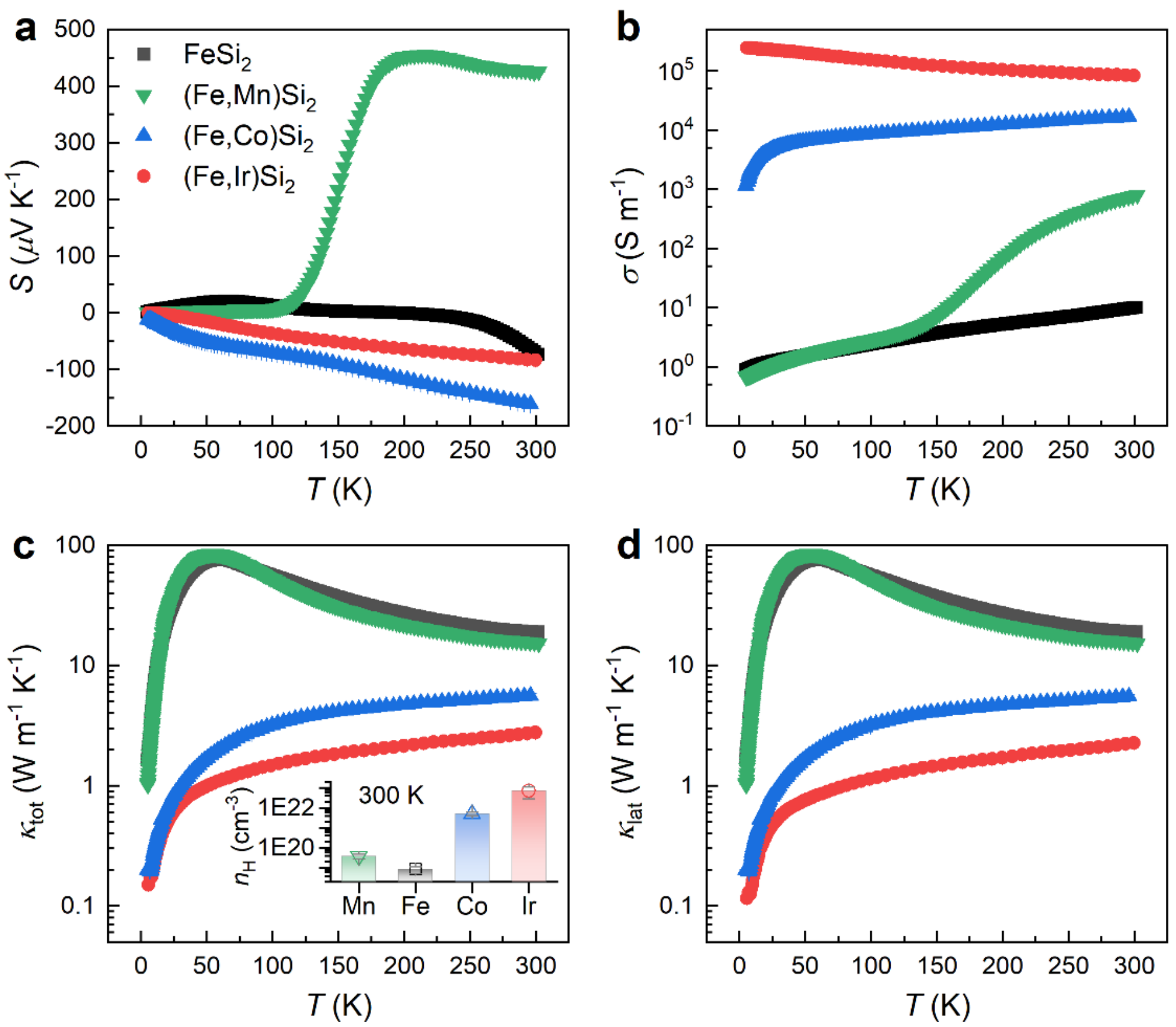

**Fig. 1 | Temperature-dependent thermoelectric transport properties of pristine and doped β-FeSi₂. a**, Seebeck coefficient, $S$. **b**, electronic conductivity, $\sigma$. The Co- and Ir-doped samples show distinct metallic behavior with high conductivity. **c**, Total thermal conductivity, $\kappa_{\mathrm{tot}}$. Inset shows the Hall carrier concentration $n_{\mathrm{H}}$ at 300 K. **d**, Lattice thermal conductivity, $\kappa_{\mathrm{lat}}$. While Mn doping retains the crystalline peak, Co and Ir doping significantly suppress $\kappa_{\mathrm{lat}}$ and exhibit a glass-like temperature dependence.

### 2.2 Semiclassical Analysis of Phonon Scattering Mechanisms

To assess whether conventional phonon scattering mechanisms can explain the observed transport behaviors, we analyzed the $\kappa_{\mathrm{lat}}$ data using the semiclassical Debye-Callaway model[46,47]. This model incorporates contributions from various scattering processes, including Umklapp (U), grain boundary (GB), point defect (PD), and electron-phonon (EP) scattering processes (refer to Supplementary Note 1 for details). Our analysis reveals a distinct disparity in phonon scattering mechanisms across different dopant types.

For pristine β-$FeSi_2$ (Fig. 2a) and p-type (Fe,Mn)$Si_2$ (Fig. 2b), phonon transport is primarily governed by intrinsic phonon-phonon Umklapp and grain boundary scattering (U+GB). Specifically, the U+GB terms alone are sufficient to reproduce the experimental data above 100 K. EP scattering becomes relevant only at low temperatures (< 100 K), likely due to carrier freeze-out or specific band-edge features in the p-type regime[37]. Consistent with the similar atomic mass and radius of Mn (54.94 u, ~1.27 Å) relative to Fe (55.85 u, ~1.26 Å), the PD scattering contribution remains negligible in (Fe,Mn)$Si_2$. Consequently, the minimal mass and strain fluctuations introduced by Mn substitution result in a $\kappa_{\mathrm{lat}}$ profile that largely mirrors that of the pristine matrix across the entire temperature range, with EP interaction playing a minor role restricted to low temperatures.

A fundamental shift occurs in the n-type systems, (Fe,Co)$Si_2$ and (Fe,Ir)$Si_2$. Here, EP scattering is no longer confined to low temperature regime but exerts a strong influence on $\kappa_{\mathrm{lat}}$ across the entire temperature range, indicating intense interaction between high-density electrons and lattice vibrations. Crucially, for the Co-doped sample, this enhanced EP scattering, combined with the U+GB background, effectively fits the data without invoking PD scattering (Fig. 2c). This is expected given the similarity between Co (58.93 u, ~1.25 Å) and Fe. In contrast, the (Fe,Ir)$Si_2$ sample necessitates both a strong EP term and a substantial PD term (Fig. 2d) to account for the large mass and radius mismatch between Ir (192.22 u, ~1.36 Å) and Fe.

However, the significantly enhanced EP term required to fit the the n-type samples raises a critical question beyond simple scattering rates. Does this strong coupling merely limit phonon lifetimes ($\tau$), or does it also induce fundamental changes to the phonon spectrum itself? Strong electron-phonon coupling can theoretically drive phonon frequency renormalization (softening)[48]. Unfortunately, the standard Debye-Callaway model assumes a static phonon dispersion relation and cannot capture such spectral modifications[49]. Consequently, the anomalously large C parameter obtained for the Co-doped sample (Supplementary Table 1) could not be interpreted as a direct measurement of the electron-phonon coupling strength. Instead, it suggests that the conventional static-lattice scattering model is insufficient to describe heat transport in this system, and there are additional, unconventional mechanisms to suppress $\kappa_{\mathrm{lat}}$. If the lattice is indeed softened, it would inherently lower phonon group velocities and the Debye temperature, acting synergistically with scattering to suppress $\kappa_{\mathrm{lat}}$. To determine whether "lattice softening" plays a covert role beyond the "enhanced scattering" picture, we turn to a direct investigation of the intrinsic lattice properties. In the following sections, we systematically probe the structural evolution, thermodynamic stiffness, and vibrational modes.

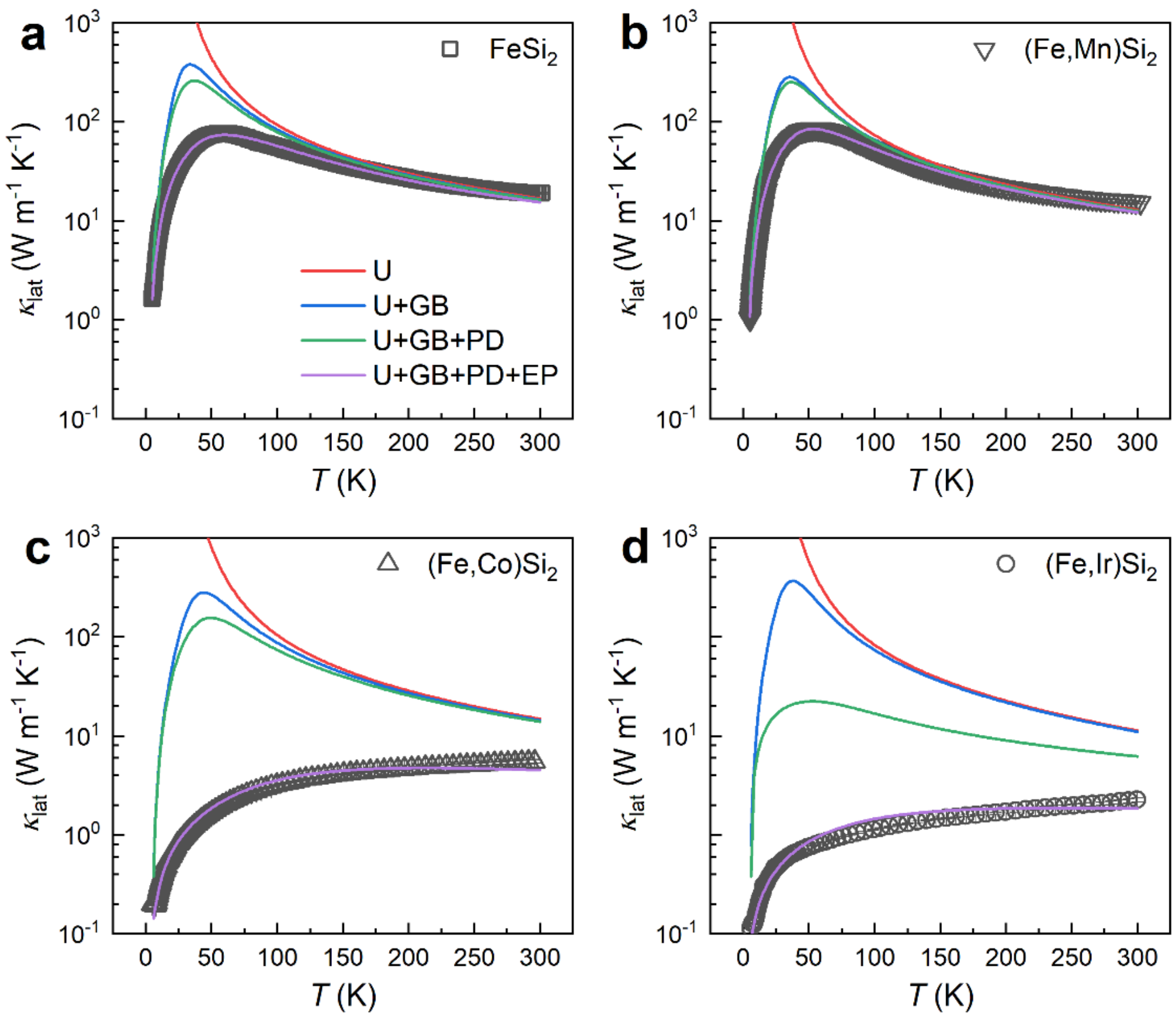

**Fig. 2 | Analysis of phonon scattering mechanisms using the Debye-Callaway model. a-d**, Experimental (symbols) and fitted curves (lines) for (a) pristine $FeSi_2$, (b) $(Fe,Mn)Si_2$, (c) $(Fe,Co)Si_2$, and (d) $(Fe,Ir)Si_2$. The contributions from Umklapp (U), grain boundary (GB), point defect (PD), and electron-phonon (EP) scattering are decoupled. Note that significant EP scattering is required across the entire temperature range for n-type samples (Co, Ir), whereas PD scattering is dominant only in the Ir-doped sample due to large mass and radius contrast. The fitting parameters are summarized in Supplementary Table 1.

### 2.3 Structural Evolution and Anomalous Lattice Expansion

To investigate the atomic-scale origin of the thermal transport behavior, we performed high-resolution neutron powder diffraction measurements at 300 K. Rietveld refinement (Fig. 3a-d) confirms that all samples are single-phase orthorhombic β-$FeSi_2$ (space group *Cmce*)[50] , with no detectable impurity phases. The β-phase unit cell (Fig. 3e) contains 48 atoms, with Fe occupying two distinct crystallographic sites (8*d* and 8*f*) within distorted Si polyhedra. This low-symmetry environment is inherently sensitive to dopant-induced local distortions[51].

The evolution of the unit cell volume ($V$), derived from Rietveld refinement of the NPD data, reveals distinct trends across different dopant types (Fig. 3f). For $(Fe,Mn)Si_2$ and $(Fe,Ir)Si_2$, the lattice expands relative to the pristine $FeSi_2$, following Vegard's law as the larger Mn and Ir replace Fe. Strikingly, Co doping also expands the lattice, despite Co having a slightly smaller atomic radius than Fe. This unexpected expansion deviates from the geometric size effect, suggesting an electronic origin rooted in lattice stiffness modulation.

This hypothesis is supported by analogous phenomena reported in vacancy-defective SrCuSb systems, where "lattice softening" led to unexpected lattice expansion[46]. Furthermore, recent theoretical calculations on Co-doped β-$FeSi_2$ predict that Co substitution disrupts the local chemical environment and reduces average interatomic force constants[52]. Physically, redunced lattice stiffness correspond to a shallower potential well and, via lattice anharmonicity, a larger equilibrium interatomic distance[53,54]. Consequently, the anomalous expansion observed here serves as direct structural evidence for lattice softening. This strongly suggests that while Ir lowers thermal conductivity through both mass fluctuation and softening, Co achieves suppression primarily through this intrinsic lattice softening, effectively compensating for its lack of mass contrast.

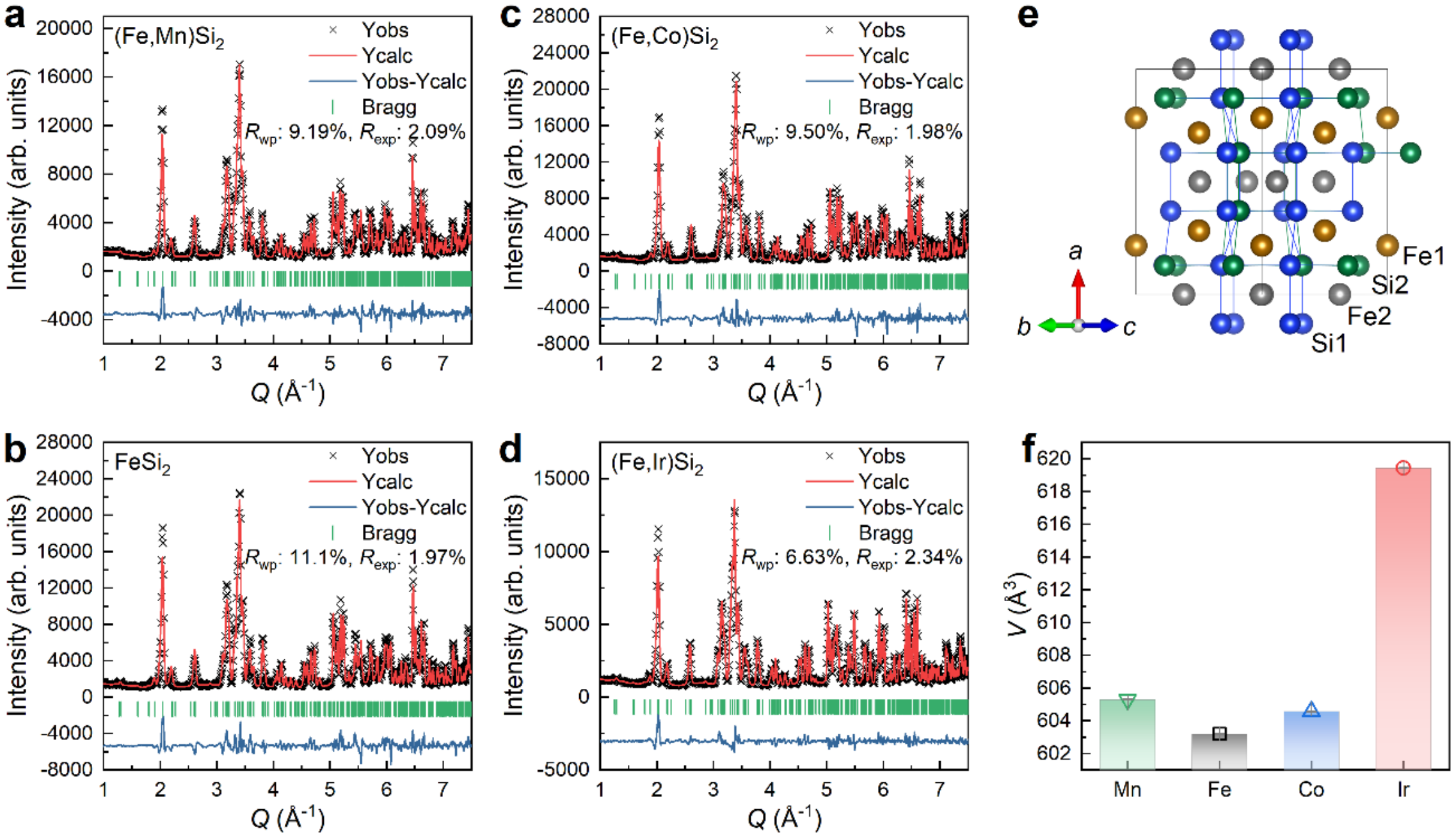

**Fig. 3 | Structural evolution of β-$FeSi_2$-based compounds. a-d**, Rietveld refinement profiles of Neutron Powder Diffraction (NPD) data at 300 K for (a) (Fe,Mn)$Si_2$, (b) pristine $FeSi_2$, (c) (Fe,Co)$Si_2$, and (d) (Fe,Ir)$Si_2$. The experimental data (black crosses), calculated profiles (red lines), and difference curves (green lines) indicate excellent fits. **e**, Schematic crystal structure of β-$FeSi_2$ projected along the [011] direction. **f**, Comparison of unit cell volumes derived from Rietveld refinements. Note the anomalous lattice expansion in the Co-doped sample despite the smaller atomic radius of Co compared to Fe, indicating a reduction in lattice stiffness.

### 2.4 Thermodynamic Stiffness and Phonon Softening

To assess the macroscopic lattice stiffness and the impact of doping on phonon populations, we measured the specific heat capacity ($C_P$) from 2 K to 300 K. Low-temperature $C_P$ data are sensitive to low-frequency acoustic phonon modes (dictating sound velocity) and low-lying optical modes that scatter heat carriers. The $C_P/T^3$ vs $T$ plots (Fig. 4) reveal a "Boson peak" feature, indicating a deviation from the Debye $T^3$ law due to the presence of low-energy optical modes[55,56]. The data were modeled using a combined Debye-Einstein model (details in Supplementary Note 2). This analysis yields characteristic Debye and Einstein temperatures, summarized in Supplementary Table 3, whose physical implications are discussed below.

The Debye temperature ($\Theta_{\mathrm{D}}$), a key indicator of global lattice stiffness, represents the cut-off frequency of acoustic phonons and serves as a proxy for the average sound velocity ($v_{\mathrm{s}}$) via $\Theta_{\mathrm{D}} \propto v_{\mathrm{s}} \propto \sqrt{k/M}$, where $k$ is the effective interatomic force constant and $M$ is the atomic mass. For pristine β-$FeSi_2$, $\Theta_{\mathrm{D}}$ was determined as 468.4 K. Doping induces a systematic softening: $\Theta_{\mathrm{D}}$ drops to 439.1 K for Mn, 418.4 K for Co, and 352.9 K for Ir. This suppression provides strong thermodynamic evidence for acoustic phonon softening. Importantly, for Mn and Co doping, given their atomic masses are nearly identical to Fe, the $\Theta_{\mathrm{D}}$ reduction must stem from a decrease in $k$. In the Ir-doped sample, the substantial mass increase works synergistically with the reduced lattice stiffness to yield the lowest $\Theta_{\mathrm{D}}$, implying the slowest phonon group velocities.

The Einstein terms are phenomenological parameters and presents the characteristic temperatures of the optical phonon DOS in different frequency ranges. They do not represent specific phonon modes, but rather the center-of-gravity of the related optical DOS. Therefore, the doping effects are clearly demonstrated by both frequency shifts and changes in the DOS shape. The Einstein temperatures ($\Theta_{\mathrm{E1}}$ and $\Theta_{\mathrm{E2}}$), representing low-lying optical modes, reveal complex local dynamics. The lower-energy Einstein modes ($\Theta_{\mathrm{E1}}$) stiffen (shift to higher frequency) for all doped samples (367.2 K for Mn, 336.4 K for Co, and 397.7 K for Ir) relative to the pristine host (317.3 K), which likely reflects a redistribution of phonon density of states (DOS) rather than a stiffening of specific bonds. Doping may introduce local strain fields that broaden the DOS or shift spectral weight to higher frequencies in certain parts of the Brillouin zone, which the simplified Debye-Einstein model captures as an effective 'hardening' parameter.

On the other hand, the relatively higher-energy optical modes ($\Theta_{\mathrm{E2}}$) exhibit divergent behavior: it hardens for (Fe,Mn)$Si_2$ (582.5 K → 617.1 K) while it softens for (Fe,Co)$Si_2$ (567.7 K) and (Fe,Ir)$Si_2$ (551.6 K). This divergence in optical phonon behavior correlates strongly with the thermal transport data. For (Fe,Mn)$Si_2$, the hardening of optical modes and the relatively high $\Theta_{\mathrm{D}}$ imply a stiffer lattice environment, explaining why $\kappa_{\mathrm{lat}}$ is only marginally suppressed at room temperature and even exceeds the pristine sample below 100 K. The lack of significant lattice softening means phonon propagation is barely impeded. In contrast, for Co- and Ir-doped samples, the softening of the $\Theta_{\mathrm{E2}}$ mode, combined with the substantial drop in $\Theta_{\mathrm{D}}$, indicates a comprehensive softening of both acoustic and high-frequency optical modes. Such global softening increases the scattering phase space for acoustic phonons via three- or four-phonon channels. This mechanism, driven by decrease in lattice stiffness (Co) and synergistic mass/strain effects (Ir), effectively suppresses the phonon group velocity across the spectrum, resulting in the pronounced $\kappa_{\mathrm{lat}}$ reduction in the n-type samples.

The thermodynamic evidence for lattice softening established above could address a critical question: to what extent does the observed reduction in lattice softening account for the magnitude of $\kappa_{\mathrm{lat}}$ suppression? We approach this through a semi-empirical analysis based on the Debye model and kinetic theory.

According to $\Theta_{\mathrm{D}} \propto v_s$, if the phonon mean-free-path remains constant, $\kappa_{\mathrm{lat}} \propto v_s^3$. The $v_s$ reduction would account for only part of the $\kappa_{\mathrm{lat}}$ suppression; any additional reduction must originate from a shortening of the phonon mean free path, i.e., enhanced scattering. Supplementary Table 4 presents this decomposition for all three doped samples at 300 K.

For (Fe,Mn)$Si_2$, the predicted $\kappa_{\mathrm{lat}}$ reduction from velocity effects alone is ~17.6%, which already accounts for most of the total observed reduction (~19.9%), confirming that Mn doping induces only marginal softening with the negligible enhancement of anharmonic scattering. In contrast, for (Fe,Co)$Si_2$, the enhanced scattering contribution dominates. This is a characteristic of lattice weakening. For (Fe,Ir)$Si_2$, the $\kappa_{\mathrm{lat}}$ reduction is a result of both the increased mass of Ir and lattice softening.

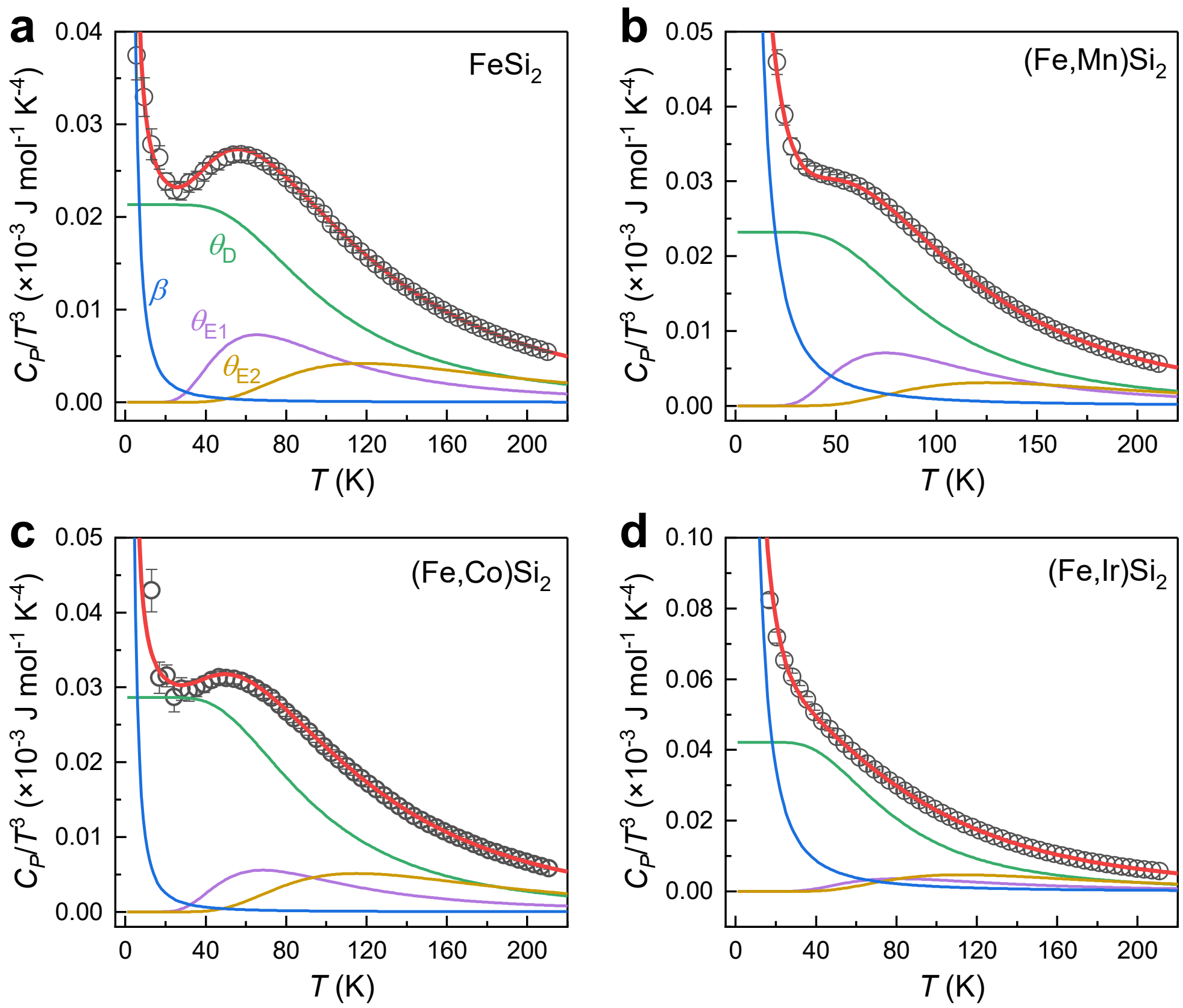

**Fig. 4 | Low-temperature heat capacity and lattice stiffness analysis. a-d**, Specific heat capacity plotted as $C_P/T^3$ vs $T$ for all samples. The open circles represent experimental data, and the red solid lines show the best fits using the combined Debye-Einstein model. The individual contributions from the Debye acoustic mode (green) and Einstein optical modes ($\Theta_{\mathrm{E1}}$ in purple, $\Theta_{\mathrm{E2}}$ in orange) are deconvoluted. The electronic contribution $\beta$ is shown in blue. The shifting of the peaks to lower temperatures and the systematic reduction of the Debye temperature ($\Theta_{\mathrm{D}}$) upon doping provide thermodynamic evidence for lattice softening.

### 2.5 Lattice Dynamics Probed by Raman Spectroscopy

To obtain direct microscopic evidence for the lattice dynamical modifications suggested by our structural and thermodynamic data, we performed room-temperature Raman spectroscopy. Fig. 5a (left panel) displays the experimental Raman spectra, which feature two prominent peaks at ~193 $cm^{-1}$ (Peak 1) and ~246 $cm^{-1}$ (Peak 2). To identify their vibrational origins, we analyzed the phonon dispersion relations, partial density of states (PDOS), and the atomic vibration patterns (Fig. 5a middle/right, 5d, 5e) for pristine β-$FeSi_2$. These two major vibration modes are assigned to the $A_g$ symmetry, involving the radial "breathing-like" motion of Fe atoms within the surrounding Si cage[42,57]. Their sensitivity to Fe-site vibrations makes them ideal local probes for dopants substituting the Fe sublattice[58].

For p-type (Fe,Mn)$Si_2$, the Raman spectrum is nearly identical to that of the pristine host in both peak position and linewidth (Fig. 5b,c). This spectroscopic stability aligns well with the preserved lattice stiffness ($\Theta_{\mathrm{D}}$) and normal lattice expansion observed earlier. It confirms that Mn substitution exerts minimal perturbation on the lattice stiffness, explaining the relatively high lattice thermal conductivity in (Fe,Mn)$Si_2$.

In contrast, n-type doping (Co, Ir) induces profound spectral changes, characterized by a pronounced redshift and significant peak broadening. First, regarding the frequency shift, the main optical modes in Co- and Ir-doped samples shift systematically to lower wavenumbers (Fig. 5b). Crucially, for (Fe,Co)$Si_2$, given the negligible mass difference between Co and Fe, this redshift provides a direct spectroscopic fingerprint of reduced interatomic force constants $k$. This corroborates the " lattice softening" mechanism inferred from the anomalous lattice expansion and suppressed Debye temperature. For (Fe,Ir)$Si_2$, the large atomic mass of Ir amplifies this softening, resulting in the largest observed redshift.

Second, the Raman peaks show substantial broadening (increased Full Width at Half Maximum, FWHM, Fig. 5c). Since the phonon linewidth $\Gamma$ is inversely proportional to the phonon lifetime $\tau$ ($\Gamma \propto 1/\tau$), this broadening signals a reduction in phonon lifetimes. In the Ir-doped sample, this is partly driven by strong point-defect scattering due to the large mass and size contrast. However, in the Co-doped sample where point defects are minimal, the broadening implies deeper dynamical origins: (1) intensified electron-phonon coupling, inherent to these high-carrier-concentration n-type systems[59]; and (2) enhanced lattice anharmonicity, a direct consequence of "lattice softening", which creates a shallower interatomic potential well and increases phonon-phonon Umklapp scattering rates[60-62].

Together, the Raman results provide a unified microscopic picture for the ultralow $\kappa_{\mathrm{lat}}$ in Co- and Ir-doped β-$FeSi_2$: "lattice softening" reduces the phonon group velocity (evidenced by redshift), while enhanced scattering (via anharmonicity and EP coupling) limits the phonon mean free path (evidenced by broadening). This dual mechanism effectively impedes heat transport, validating the superior thermoelectric performance observed in these systems.

Although Mn- and Co-elements have close atomic masses and radii, their effects on lattice stiffness are quite different. Co-doping (n-type) introduces an extra valence electron populates Fe-Si anti-bonding states [52], and effectively reduces the interatomic force constants, providing a microscopic explanation for the anomalous lattice expansion and the $\Theta_{\mathrm{D}}$ reduction observed in $(Fe,Co)Si_2$. On the other side, Mn-doping (p-type) removes an electron from the system. Therefore, the lattice rigidity of $(Fe,Mn)Si_2$ is clearly affected by the bonding states, which is confirmed by both minimum Raman redshift and $\Theta_{\mathrm{D}}$ reduction.

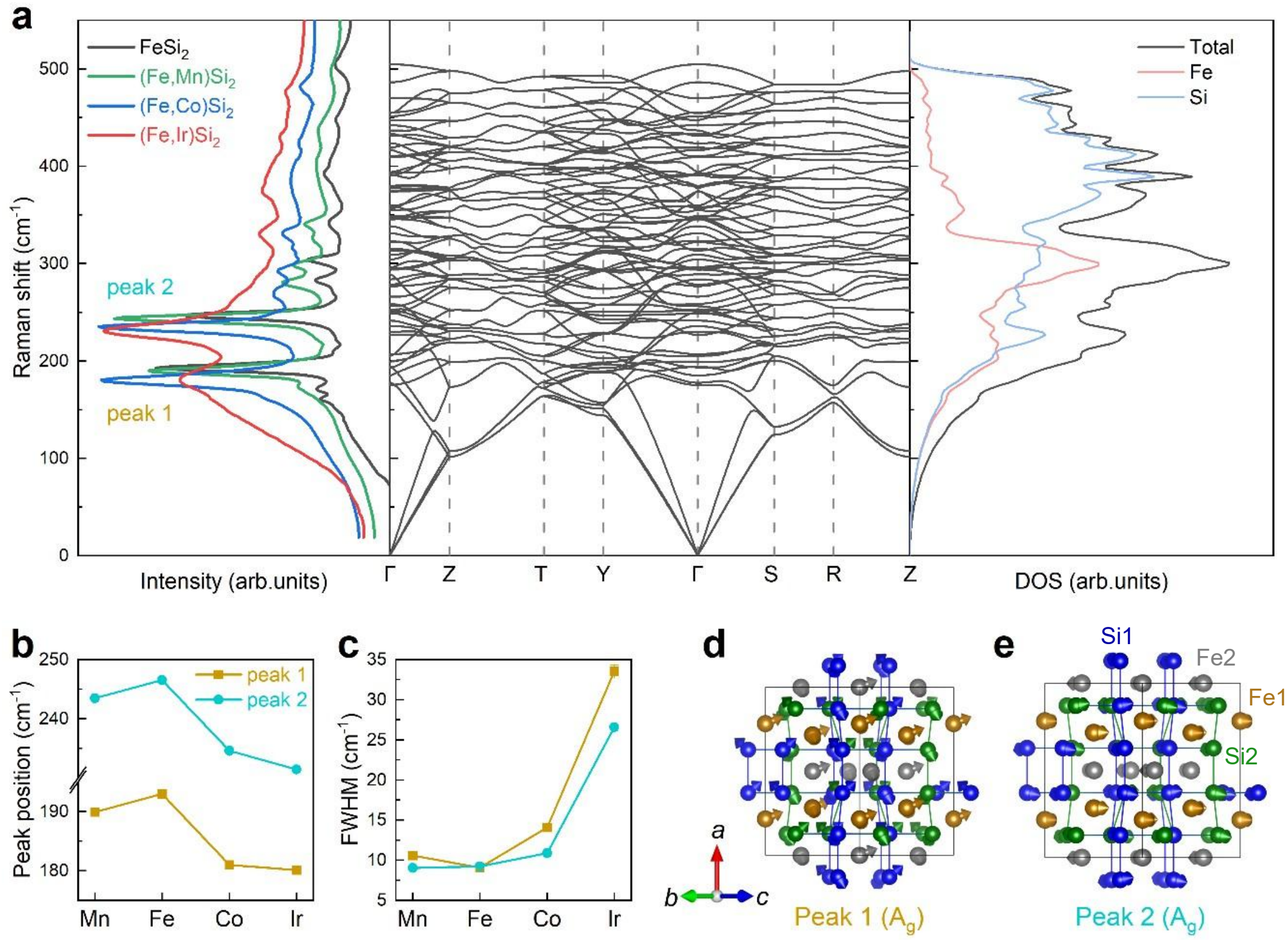

**Fig. 5 | Lattice dynamics probed by Raman spectroscopy.** **a**, Experimental Raman spectra of pristine and doped β -$FeSi_2$ samples at room temperature (left panel) aligned with the theoretical phonon dispersion and density of states (DOS) for pristine β-$FeSi_2$ (middle and right panels). Two characteristic $A_g$ modes involving Fe-atom vibrations are highlighted as Peak 1 and Peak 2. **b**,**c**, The frequency shift (b) and Full Width at Half Maximum (FWHM, c) of the two characteristic Raman peaks. While the p-type Mn-doped sample shows minimal changes, the n-type Co- and Ir-doped samples exhibit significant redshift (softening) and broadening. The anomalous softening and broadening in the Co-doped sample, despite the lack of mass contrast, provide direct evidence for weakening lattice stiffness and enhanced phonon scattering. **d**,**e**, Schematic representation of the atomic vibration patterns for these two $A_g$ optical modes.

## 3. Conclusions

In summary, this work provides a compelling atomistic link between dopant-induced lattice stiffness modifications and macroscopic thermal transport in β-$FeSi_2$. While conventional strategies rely heavily on mass and strain contrasts, we demonstrate that doping-enhanced lattice softening, manifesting as phonon frequency renormalization, serves as a distinct and potent mechanism for $\kappa_{\mathrm{lat}}$ suppression. Most notably, the substantial ~71% reduction in $\kappa_{\mathrm{lat}}$ of Co-doped β-$FeSi_2$, achieved in the absence of significant mass or size mismatch, provides robust evidence that lattice stiffness modulation is a highly effective pathway for thermal management. These findings provide a new design paradigm for thermoelectrics: beyond simply seeking heavy-element dopants, targeting agents that induce lattice softening offers a direct and effective pathway to ultralow lattice thermal conductivity by modifying intrinsic lattice stiffness.

## 4. Experimental section

**Sample Preparation.** High-purity elements, Fe (Alfa Aesar, 99.98%, granules), Ir (Alfa Aesar, 99.99%, powders), Co (Alfa Aesar, 99.99%, powders), Mn (Alfa Aesar, 99.98%, granules), and Si (Alfa Aesar, 99.98%, granules), were weighed out in the atomic ratio of pristine β-$FeSi_2$ and doped $Fe_{0.84}Ir_{0.16}Si_2$, $Fe_{0.94}Co_{0.06}Si_2$, and $Fe_{0.92}Mn_{0.08}Si_2$. The doping concentrations were selected based on previously reported optimized thermoelectric performance[32]. For simplicity, the pristine and doped samples will be hereafter referred to as $FeSi_2$, $(Fe,Ir)Si_2$, $(Fe,Co)Si_2$, and $(Fe,Mn)Si_2$, respectively. The raw materials were arc-melted in an argon atmosphere. The melted ingots were manually ground into fine powders. Then, the fine powders were sintered at 1223 K by SPS (Sumitomo, SPS-2040). The sintering pressure was 65 MPa and the holding time was 10 min. The sintered products were sealed in silica tubes under vacuum and annealed at 1423 K for 1 h, and then lowered to 1173 K for another 48-h annealing, following by natural cooling to room temperature.

**Structural characterization.** Room-temperature X-ray diffraction measurements were performed using an in-house Rigaku diffractometer to check the sample quality. High-resolution neutron powder diffraction (NPD) data were collected using the Echidna powder diffractometer at the Australian Nuclear Science and Technology Organisation (ANSTO)[63]. Rietveld refinements for NPD patterns were conducted using FullProf suite[64].

**Physical properties measurements.** Heat capacity measurements were conducted from 2 to 300 K using the Heat Capacity Option of the Physical Property Measurement System (PPMS, Quantum Design). The total thermal conductivity ($\kappa_{\mathrm{tot}}$) was measured using the Thermal Transport Options (TTO) of PPMS with the four-probe lead configuration. The lattice thermal conductivity ($\kappa_{\mathrm{lat}}$) was then calculated by subtracting the electronic contribution ($\kappa_{\mathrm{ele}}$) from $\kappa_{\mathrm{tot}}$. obtained using $\kappa_{\mathrm{lat}} = \kappa_{\mathrm{tot}} - \kappa_{\mathrm{ele}}$. $\kappa_{\mathrm{ele}}$ was estimated using the Wiedemann-Franz law, $\kappa_{\mathrm{ele}} = L\sigma T$, where $L$ is the Lorentz number. $L$ was calculated using the empirical formula $L = 1.5 + \exp(-|S|/116)$, with $L$ and $S$ in units of $10^{-8}$ W $K^{-2}$ and μV $K^{-1}$, respectively[65].

**Raman spectra.** Room-temperature Raman spectroscopy was performed on pristine and doped samples to probe the phonon vibrational modes. The spectra were acquired using a DXR Raman Microscope equipped with a 532.1 nm excitation laser, covering a spectral range of 50-600 $cm^{-1}$.

## Declarations

### Availability of data and material

Data of this study are available from the corresponding author upon reasonable request.

### Funding information

This work is supported by the National Key Research and Development Program of China (no. 2024YFA1409202, to J.M., 2024YFE0110005, to Q.R.), the National Natural Science Foundation of China (no. 12474024, to Q.R.; no. 52101236, to Q.R.; no. 12334008 to J.M.), Guangdong Basic and Applied Basic Research Foundation (no. 2024A1515140063, to Q.R.), Guangdong Provincial Key Laboratory of Extreme Conditions (no. 2023B1212010002, to M.S.).

### Author contributions

Q.R., P.Q., and J.M. proposed and supervised the project. P.Q. and C.C. synthesized the samples. C.Z. performed thermal characterization with the assistance from W.L., J.S., F.C., W.X., Y.L., and M.S. The XRD measurements were performed by C.Z. with help from S.H. and G.W. The high-resolution NPD were carried out by J.M. and C.W.W., and the Rietveld refinements were done by C.Z. and Q.R. The Raman spectra were collected by X.H and analyzed by C.Z. and Q.R. The theoretical calculations were conducted by Y.C and S.D.. C.Z. and Q.R. organized the data and drafted the manuscript. All authors contributed helpful discussions to this work.

### Competing interests

The authors declare no conflict of interest.

## Supplementary Information

# Dopant-modulated lattice stiffness and drastic thermal conductivity reduction in β-$FeSi_2$ thermoelectrics

### Supplementary Note 1. Debye-Callaway Fitting

In order to obtain different contributions of the lattice thermal conductivity, and the nonmagnetic simplified Debye-Callaway model based on Boltzmann distribution was applied[1]:

$$\kappa_{\mathrm{lat}} = \frac{k_{\mathrm{B}}}{2\pi v^2}\left(\frac{k_{\mathrm{B}}T}{\hbar}\right)^3 \int_0^{\frac{\Theta_{\mathrm{D}}}{T}} \tau \frac{x^4 e^x}{(e^x-1)^2} dx \tag{S1}$$

$$\tau^{-1} = \frac{v}{L} + A\omega^4 + B\omega^2 T e^{-\frac{\Theta_{\mathrm{D}}}{3T}} + C\omega^2 \tag{S2}$$

where $x = \hbar\omega/k_{\mathrm{B}}T$ is the reduced phonon energy, $\omega$ is the phonon frequency, $\hbar$ is the reduced Plank constant, $k_{\mathrm{B}}$ is the Boltzmann constant and $\nu$ is the average sound velocity. $\Theta_{\mathrm{D}}$ denote the Debye temperature. $\tau$ is the reduced phonon lifetime, which affected by the boundary scattering ($\nu/L$), point-defect (PD) scattering ($A\omega^4$), Umklapp (U) process ($B\omega^2 T e^{-\Theta_{\mathrm{D}}/3T}$), and electron-phonon (EP) scattering ($C\omega^2$). Where $L$ is the grain size for the polycrystal, A, B and C are fitting parameters. The results are summarized in Supplementary Table. 1.

### Supplementary Note 2. Heat Capacity

We use the Debye-Einstein model to fit the experimental heat capacity, $C_P$,

$$C_P = \beta T + 9R\frac{\sum_i A_i}{(n-1)}\left(\frac{T}{\Theta_{\mathrm{D}}}\right)^3 \int_0^{\frac{\Theta_{\mathrm{D}}}{T}} \frac{\varepsilon^4 e^\varepsilon}{(e^\varepsilon-1)^2} d\varepsilon + 3R\sum_i A_i \frac{\left(\frac{\Theta_{\mathrm{Ei}}}{\mathrm{T}}\right)^2 e^{\frac{\Theta_{\mathrm{E}i}}{T}}}{\left(e^{\frac{\Theta_{\mathrm{E}i}}{T}}-1\right)^2} \tag{S3}$$

where, $\beta$ represents the electronic contribution to the specific heat. The second term gives the Debye mode contribution. The third term corresponds to the contribution from the Einstein mode ($\Theta_{\mathrm{E}i}$, $i^{\mathrm{th}}$ Einstein temperature), where $A_i$ is the pre-factor of $i^{\mathrm{th}}$ Einstein mode ( $\sum_i A_i$ should approximate the n-1 atoms in a unit cell). $R$ is the gas constant, $n$ is the number of atoms in a unit cell. The results are summarized in Supplementary Table 3.

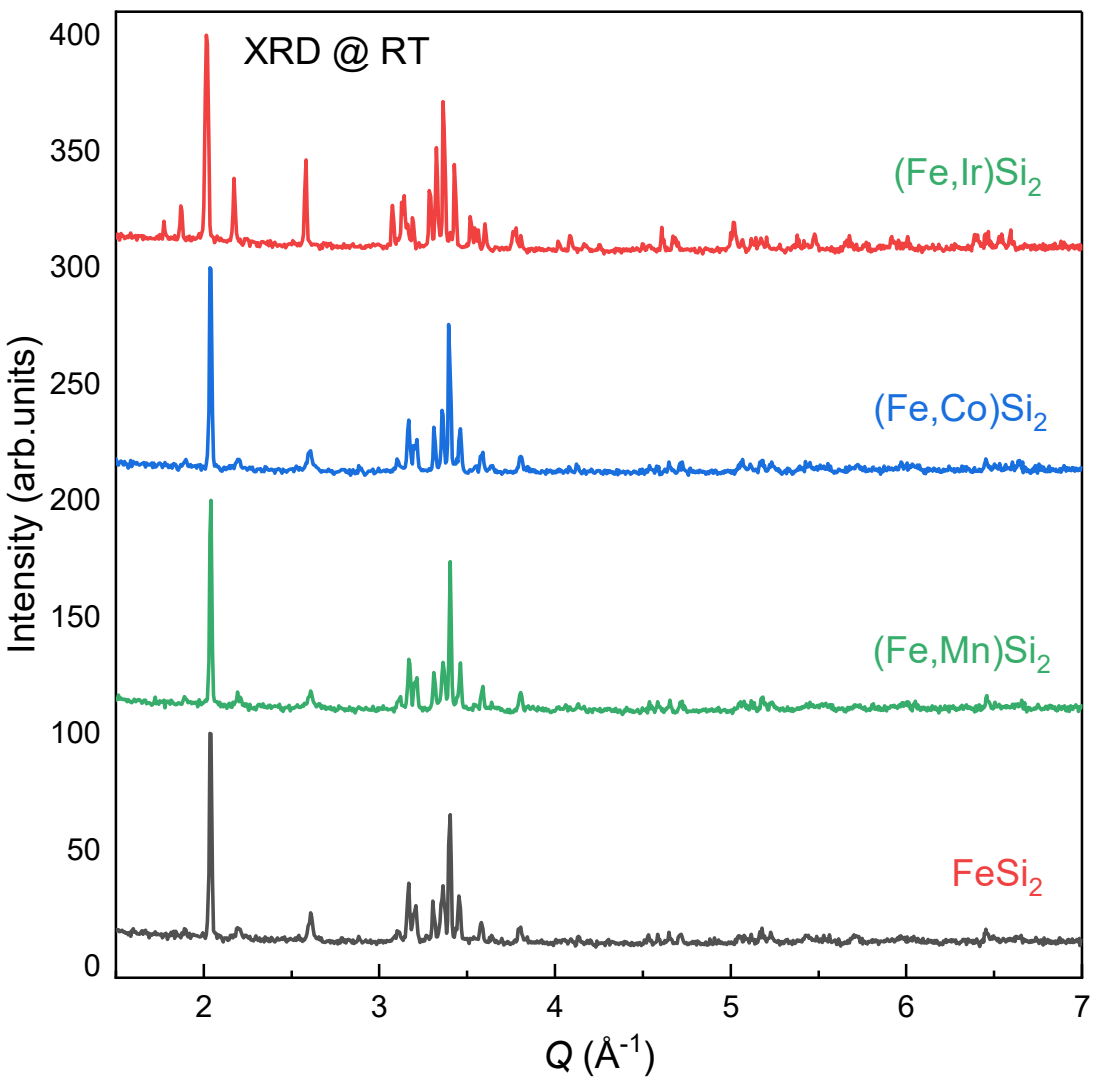

**Supplementary Fig. 1 The X-ray diffraction (XRD) patterns for the β-$FeSi_2$-based compounds.** The patterns were collected with Cu Kα radiation at room temperature (RT).

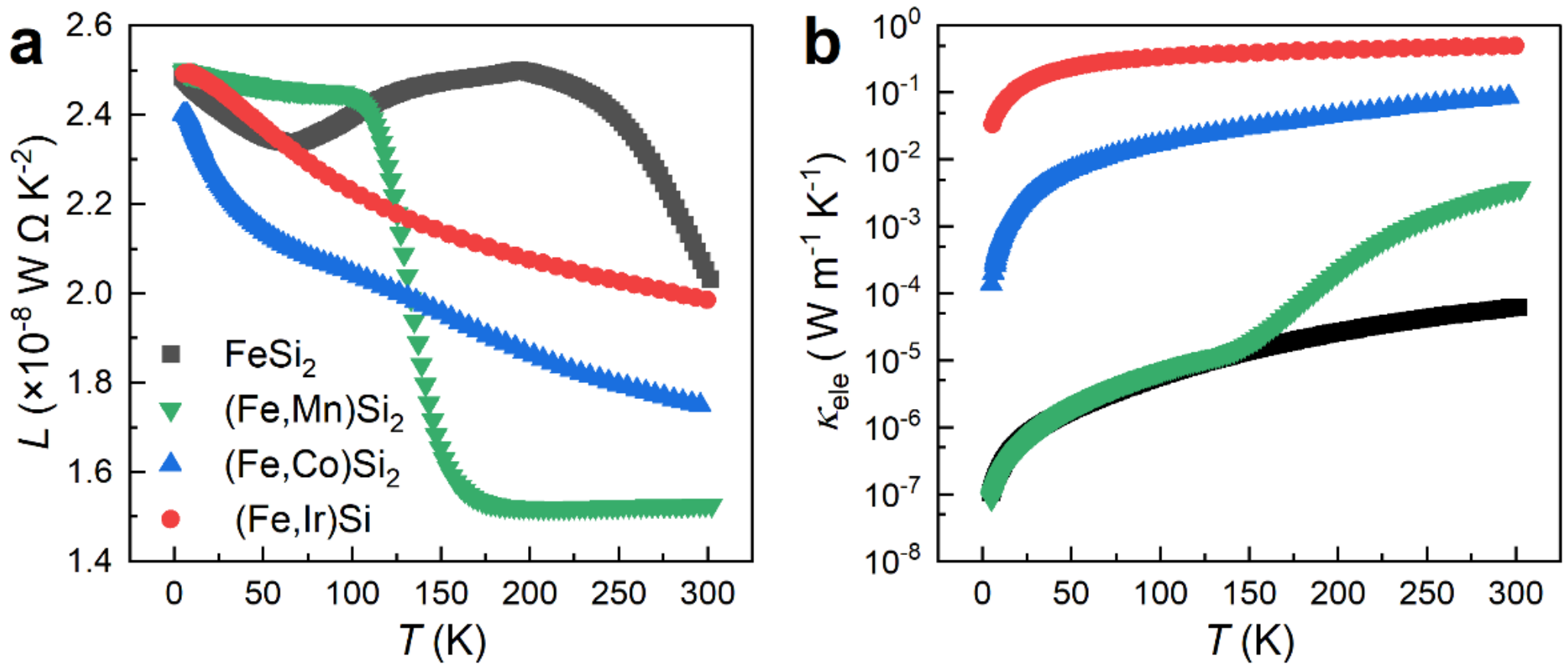

**Supplementary Fig. 2 Estimation of electronic thermal transport contributions.** **a**, Temperature dependence of the Lorenz number ($L$) calculated using the single parabolic band (SPB) model for all samples. **b**, The calculated electronic thermal conductivity ($\kappa_{\mathrm{ele}}$) as a function of temperature. $\kappa_{\mathrm{ele}}$ remains significantly lower than the total thermal conductivity for all compositions, justifying the focus on lattice thermal transport.

**Supplementary Table 1 | Parameters obtained by fitting experimental lattice thermal conductivity.** Parameters include the Debye temperature ($\Theta_{\mathrm{D}}$), average group velocity ($v$), and the pre-factors A, B, and C for the PD, U, and EP scatterings, respectively.

| Parameter | $FeSi_2$ | $(Fe,Mn)Si_2$ | $(Fe,Co)Si_2$ | $(Fe,Ir)Si_2$ |
|---|---|---|---|---|
| A ($10^{-42}$ $s^3$) | 4.251(4) | 1.751(7) | 5.574(4) | 9.779(4) |
| B ($10^{-18}$ $K^{-1}s^2$) | 1.295(4) | 1.356(9) | 1.836(4) | 1.958(4) |
| C ($10^{-15}$ s) | 1.738(4) | 2.213(9) | 5.628(4) | 9.782(4) |
| $L$ ($10^{-6}$ m) | 4.246(4) | 4.105(4) | 4.584(4) | 4.702(4) |
| $\Theta_D$ (K) | 673 | 656 | 597 | 559[2] |
| $v$ (m $s^{-1}$) | 5663[3] | 5211 | 5178[3] | 4361[2] |
| $R^2$ | 0.9653 | 0.9533 | 0.9698 | 0.9782 |

**Supplementary Table 2 | The crystallographic parameters of the $\beta$-$FeSi_2$-based compounds.**

| Parameter | $a$ (Å) | $b$ (Å) | $c$ (Å) | $V$ ($Å^3$) | $R_{wp}$ (%) |
|---|---|---|---|---|---|
| $FeSi_2$ | 9.8770(2) | 7.7996(2) | 7.8325(2) | 603.389(6) | 11.1 |
| $(Fe,Mn)Si_2$ | 9.8745(3) | 7.8089(2) | 7.8466(4) | 605.043(4) | 9.19 |
| $(Fe,Co)Si_2$ | 9.9085(2) | 7.8042(3) | 7.8426(2) | 606.451(9) | 9.50 |
| $(Fe,Ir)Si_2$ | 9.9862(2) | 7.8412(2) | 7.9097(3) | 619.363(5) | 6.63 |

**Supplementary Table 3 | Lattice dynamical parameters derived from heat capacity modeling.** Parameters include the electronic coefficient ($\beta$), Debye temperature ($\Theta_{\mathrm{D}}$), Einstein temperatures ($\Theta_{\mathrm{E1}}$, $\Theta_{\mathrm{E2}}$), and their pre-factors ($A_1$, $A_2$).

| Parameter | $FeSi_2$ | $(Fe,Mn)Si_2$ | $(Fe,Co)Si_2$ | $(Fe,Ir)Si_2$ |
|---|---|---|---|---|
| $\beta$ ($10^{-3}$ J mol $K^{-1}$) | 0.013(2) | 9.349(3) | 8.011(4) | 13.385(4) |
| $A_1$ | 0.441(7) | 0.657(7) | 0.398(7) | 1.288(7) |
| $A_2$ | 1.638(7) | 1.364(7) | 1.761(7) | 0.641(7) |
| $\Theta_{\mathrm{D}}$ (K) | 468.4(4) | 439.1(4) | 418.4(4) | 352.9(4) |
| $\Theta_{\mathrm{E1}}$ (K) | 317.3(4) | 367.2(4) | 336.4(4) | 397.7(4) |
| $\Theta_{\mathrm{E2}}$ (K) | 582.5(4) | 617.1(4) | 567.7(4) | 551.6(4) |
| $R^2$ | 0.9953 | 0.9541 | 0.9919 | 0.9865 |

**Supplementary Table 4 | Semi-empirical estimate of $\kappa_{\mathrm{lat}}$ reduction at 300 K for $\beta$-$FeSi_2$-based compounds.**

| Parameter | $\Delta v_{\mathrm{s}}$ (%) | $\kappa_{\mathrm{lat}}$ reduction (%) | Velocity contribution (%) | Scattering contribution (%) |
|---|---|---|---|---|
| $(Fe,Mn)Si_2$ | −6.3 | 19.9 | 17.6 | 2.3 |
| $(Fe,Co)Si_2$ | −10.7 | 71.2 | 28.8 | 42.4 |
| $(Fe,Ir)Si_2$ | −24.7 | 88.0 | 57.2 | 30.8 |